\documentclass{jps-cp}
\usepackage[dvipdfmx]{color}
\usepackage{txfonts} 

\title{Field-direction Dependence of Majorana-mediated Spin Transport}

\author{
  Hirokazu \textsc{Taguchi}$^1$, Akihisa \textsc{Koga}$^1$, Yuta \textsc{Murakami}$^2$,
  Joji \textsc{Nasu}$^3$, Hiroki \textsc{Tsuchiura}$^4$}

\inst{$^1$Department of Physics, Tokyo Institute of Technology, Meguro,
  Tokyo 152-8551, Japan \\
  $^2$Center for Emergent Matter Science, RIKEN, Wako 351-0198, Japan\\
  $^3$Department of Physics, Tohoku University, Sendai 980-8578, Japan\\ 
  $^4$Department of Applied Physics, Tohoku University, Sendai 980-8579, Japan 
}
\email{koga@phys.titech.ac.jp}

\recdate{June 3, 2022}

\abst{
  We study the field-direction dependence of
  the Majorana-mediated spin transport in the Kitaev clusters
  with zigzag and armchair edges, 
  applying a static magnetic field to one of the edges
  and a magnetic pulsed field to the other edges.
  By means of the exact diagonalization method,
  we calculate the time-evolution of the spin moments in both edge regions
  to clarify how the directions of two fields and shape of the edges
  affect the Majorana-mediated spin transport.
}

\kword{Majorana fermions, Kitaev model, spin trasport}

\begin{document}
\maketitle

\section{Introduction}
Transport phenomena of spin excitations have attracted considerable
attention in condensed matter physics.
In insulating magnets, spin excitations are mainly carried by magnons.
It was recently suggested that, in the nonmagnetic ground state
of the one-dimensional Heisenberg system,
the spin transport is realized by spinons,
which are elementary excitations fractionalized from spins.
This has been observed in the spin Seebeck experiments for the cuprate
$\rm Sr_2CuO_3$~\cite{SpinonSpinCurrent}.
The Kitaev quantum spin model~\cite{Kitaev_model} is another candidate of
nonmagnetic systems.
In the system, the elementary excitations are described
by two kinds of quasiparticles: itinerant and
localized Majorana fermions~\cite{Motome_Nasu_Review},
and their roles have extensively been
discussed~\cite{NasuDoublePeak,Feldmeier,Konig,Pereira,Udagawa,KogaCor,Hoon}.
It has also been clarified that
the itinerant Majorana fermions carry the spin excitations
although no magnetic moments appear in the genuine Kitaev
model~\cite{Minakawa_2020,Taguchi_2021,Taguchi2}.
Such a spin transport can be examined in the time-evolution of
the Kitaev cluster with edges;
the introduction of the magnetic pulse in one of the edges does not induce
the spin oscillations in the bulk region, but induces them
in the other edge with the static magnetic field.
In the previous studies,
this unusual phenomenon has been clarified in the cluster, where
both magnetic pulse and static fields are applied along the $S^z$ direction.
Then, a question arises: how do similar spin oscillations appear
even when the magnetic pulse are introduced along the $S^x$ direction?
Do itinerant Majorana fermions carry the direction of the magnetic field pulse
as well as the spin excitations?
This should be important to observe spin oscillations
in the realistic systems.

Motivated by this, we consider the Kitaev clusters,
where the magnetic fields in two edge regions are applied in the distinct directions.
We study the field-direction dependent spin propagation
by means of the exact diagonalization method.
The differece in the shape of edges is also addressed.

\section{Model and Method}\label{sec2}

\begin{figure}[htb]
  \centering
  \includegraphics[width=\linewidth]{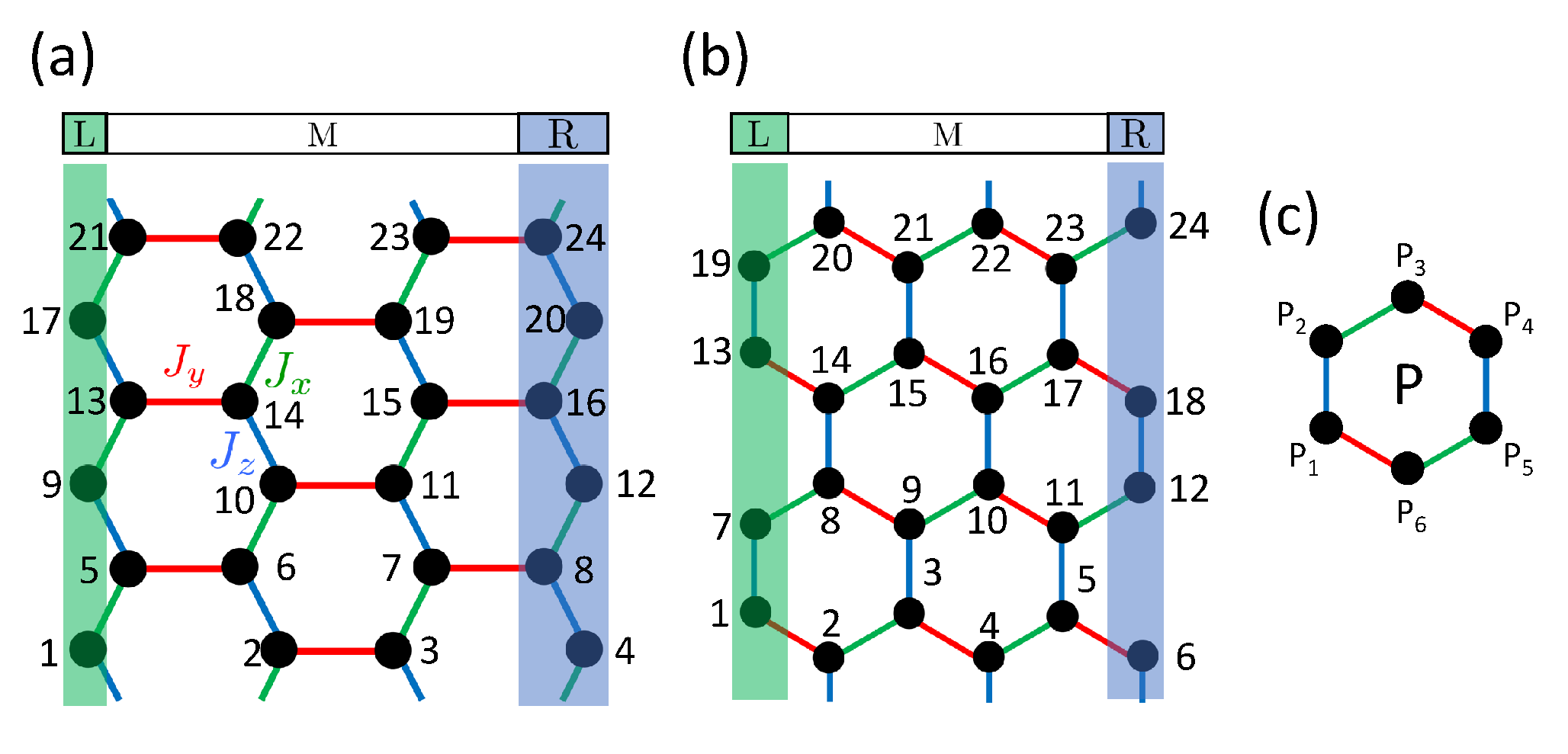}
  \caption{Kitaev clusters with (a) zigzag edges and (b) armchair edges.
    Green, red, and blue lines indicate $x$, $y$, and $z$ bonds, respectively.
    The static magnetic field $h_R$ is applied in the right (R) region, and
    no magnetic field is applied in the middle (M) region.
    A time-dependent pulsed magnetic field is introduced in the left (L) region.
    (c) Plaquette $P$ with sites marked $P_1$, $P_2$, \dots, $P_6$
    shown for the operator $W_P$.
  }
  \label{fig:system}
\end{figure}

We consider the Kitaev clusters with zigzag and armchair edges,
which are schematically shown in Figs.~\ref{fig:system}(a) and (b).
The Hamiltonian is given as
\begin{eqnarray}
  H(t) &=& -J\sum_{\langle i,j\rangle_x}S_i^x S_j^x
  -J\sum_{\langle i,j\rangle_y}S_i^y S_j^y
  -J\sum_{\langle i,j\rangle_z}S_i^z S_j^z
  -h_R\sum_{i \in R}  S^z_i
  - h_L(t)\sum_{i \in L}{\bf n}\cdot {\bf S}_i,\label{H}
\end{eqnarray}
where $\langle i, j\rangle_\mu$ indicates
the nearest-neighbor pair on the $\mu(=x,y,z)$-bonds.
The $x$-, $y$-, and $z$-bonds are shown as green, red, and blue lines
in Fig.~\ref{fig:system}.
$S_i^\mu$
is the $\mu$ component of an $S=1/2$ spin operator at site $i$.
$J$ is the exchange coupling between the nearest neighbor sites.
$h_R(=0.01J)$ represents the static magnetic field in the R region.
For simplicity, its direction is fixed as the $S^z$ direction.
In the L region, the time-dependent magnetic field is applied
in the direction ${\bf n}$ and its magnitude is given by the Gaussian form as
\begin{align}
  h_L(t) = \frac{A}{\sqrt{2\pi}\sigma}\exp{\left[\frac{t^2}{2\sigma^2} \right]},
  \label{eq:hLt}
\end{align}
where $A$ and $\sigma$ are strength and width of the pulse.
Here, we set $\sigma = 2/J$ and $A = 1$.

Now, we consider the famous local operator as 
\begin{eqnarray}
  W_p=2^6 S_{P_1}^xS_{P_2}^yS_{P_3}^zS_{P_4}^xS_{P_5}^yS_{P_6}^z,
\end{eqnarray}
where $P_i\;(i=1,2,\dots, 6)$ is the site in the plaquette $P$
[see Fig.~\ref{fig:system}(c)].
Each local operator commutes with
the Kitaev Hamiltonian eq.~(\ref{H}) with $h_R=h_L(t)=0$.
This means the existence of the local conserved quantities,
and each eigenstate of the model is classified by the subspace
with $\{w_P\}$, where $w_P$ is the eigenvalue of $W_P$.
Furthermore, it is known that, in the ground state,
$w_P=1$ for each plaquette $P$~\cite{Kitaev_model}.
This leads to the important features for the Kitaev model.
For example, one can prove that
$\langle S_{P_1}^y\rangle=\langle S_{P_1}^z\rangle=
\langle S_{P_2}^z\rangle=\langle S_{P_2}^x\rangle=\cdots=0$
for the state with $w_P=1$ on an isolated plaquette $P$,
as shown in Fig.~\ref{fig:system}(c).
By taking into account the conserved quantity for each plaqutte
in the whole honeycomb lattice,
one can prove the absence of magnetic moments at each site.
In the sence, in the finite cluster,
a certain direction in the magnetic moments may appear
since we cannot prove the absence of the moment in terms of the local conserved quantities,
{\it eg.} the $y$-component of the spin moment at site 9
in the Kitaev cluster with zigzag edges
even when $h_R=h_L(t)=0$ [see Fig.~\ref{fig:system}(a)].
Furthermore, the Zeeman terms does not commute with the local operator $W_P$,
the spin moments, in general, appear in both L and R regions
under magnetic fields.
These facts are important to observe the spin oscillations
in the Kitaev system.
In the following, we focus on the spin moments in the L and R regions.
Here, we calculate the change in the moment as
\begin{align}
  \Delta {\bf S}(t)=\langle {\bf S}(t)\rangle-\langle {\bf S}(-\infty)\rangle,
\end{align}
where ${\bf S}(t)$ is the spin moment at time $t$.

In this study, we discuss how the shape of the edges and/or
direction of the magnetic field affects the spin transport.
To this end, we treat the 24-site Kitaev clusters
with armchair and zigzag edges, as shown in Figs.~\ref{fig:system}(a) and (b).
Then, using the Lanczos method~\cite{Park_1986, Saad_1992, Druskin_1995, Marlis_1997, Hochbruck_1998, Hochbruck1999},
we examine the time evolution of the system
after the magnetic pulse is applied in the $S^x$ and $S^z$ directions.

\section{Results}\label{sec3}

\begin{figure}[htb]
  \centering
  \includegraphics[width=\linewidth]{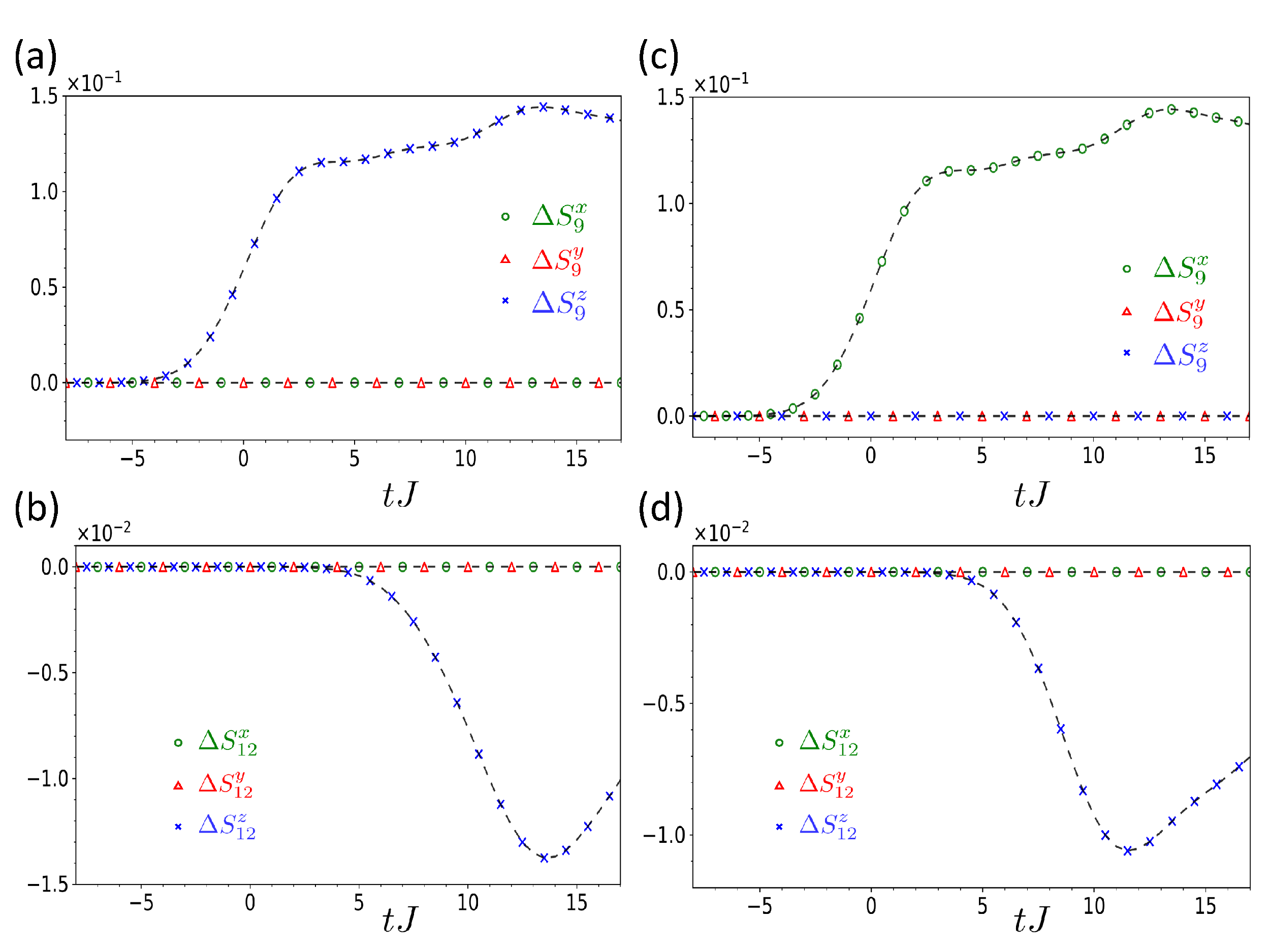}  
  \caption{Real-time evolution of the Kitaev cluster with zigzag edges.
    (a) and (b) represent the change in the spin moments in the L and R regions
    when the direction of the injected magnetic pulse is along the $S^z$ direction
    (${\bf n}={\bf z}$).
    (c) and (d) represent the change in the spin moments in the L and R regions
    when ${\bf n}={\bf x}$.
  }
  \label{fig:Zig}
\end{figure}
First, we treat the zigzag-edge Kitaev cluster
to discuss how the direction of the magnetic pulse induces
the spin oscillations in the R region
where the direction of the static magnetic field is fixed in the $S^z$ direction.
Figure~\ref{fig:Zig} shows
the time-evolution of the change in the spin moments.
When the magnetic pulse with ${\bf n}={\bf z}$ is introduced in the L region,
it is naively expected that
the spin oscillations in the $S^z$ direction appear in the L region.
In fact, the magnetic moments are induced around $t=0$, as shown in Fig.~\ref{fig:Zig}(a).
The subpeak-like structure around $t=12/J$ should
originate from complex factors 
such as the oscillation in the flux next of edges, the amplitude and width of the Gaussican pulse, finite size effect in the cluster, etc.
This is beyond the main scope of our study,
and will be discussed in the future.
In the middle region bounded by the L and R regions,
there are local conserved quantities $W_p$, and thereby
magnetic moments never appear (not shown).
In the R region, the spin oscillations are induced only in the $S^z$ direction
around $t=5/J$.
This is consistent with the fact that
the Hamiltonian is invariant under
the symmetry operation $\prod_i U_i^z$,
where $U^\mu_i(=\exp[i\pi S_i^\mu])$ is the $\pi$-rotation along the $S^\mu$ direction
for the $i$th spin.
It has been clarified that this unusual spin propagation
is mediated by the itinerant Majorana fermions,
which is one of the fractionalized particles
in the Kitaev model~\cite{Minakawa_2020}.
On the other hand, different behavior appears when ${\bf n}={\bf x}$.
In the L region, the oscillations in magnetic moments are induced 
along the applied field, as shown in Fig.~\ref{fig:Zig}(c).
Namely, the curves of $z$ component in (a) and of $x$ component in (c)
are the same due to the symmetry of the system.
On the other hand, in the R region,
the magnetic moments are induced along the $S^z$ direction.
In this case, the direction of the spin oscillations is not related to
the magnetic pulse introduced in the L region but the static field in the R region.
This can be explained by the following symmetry argument.
The Kitaev system can be regarded as the linked zigzag chains parallel to the edges.
The Hamiltonian with ${\bf n}={\bf x}$ is invariant
under the symmetry operation $\prod_{i\in o}U_i^x\prod_{i\in e}U_i^z$,
where $o$ ($e$) represents the set of spins
in odd (even) numbered zigzag chains from the left edge.
Therefore, we can conclude the absence of the $x$ and $y$ components
in the induced moment in the R region. 
This may suggest that the itinerant Majorana fermions propagate
the wave packet triggered
by the magnetic pulse, while cannot propagate its direction. 

We also consider the Kitaev cluster with armchair edges.
\begin{figure}[htb]
  \centering
  \includegraphics[width=\linewidth]{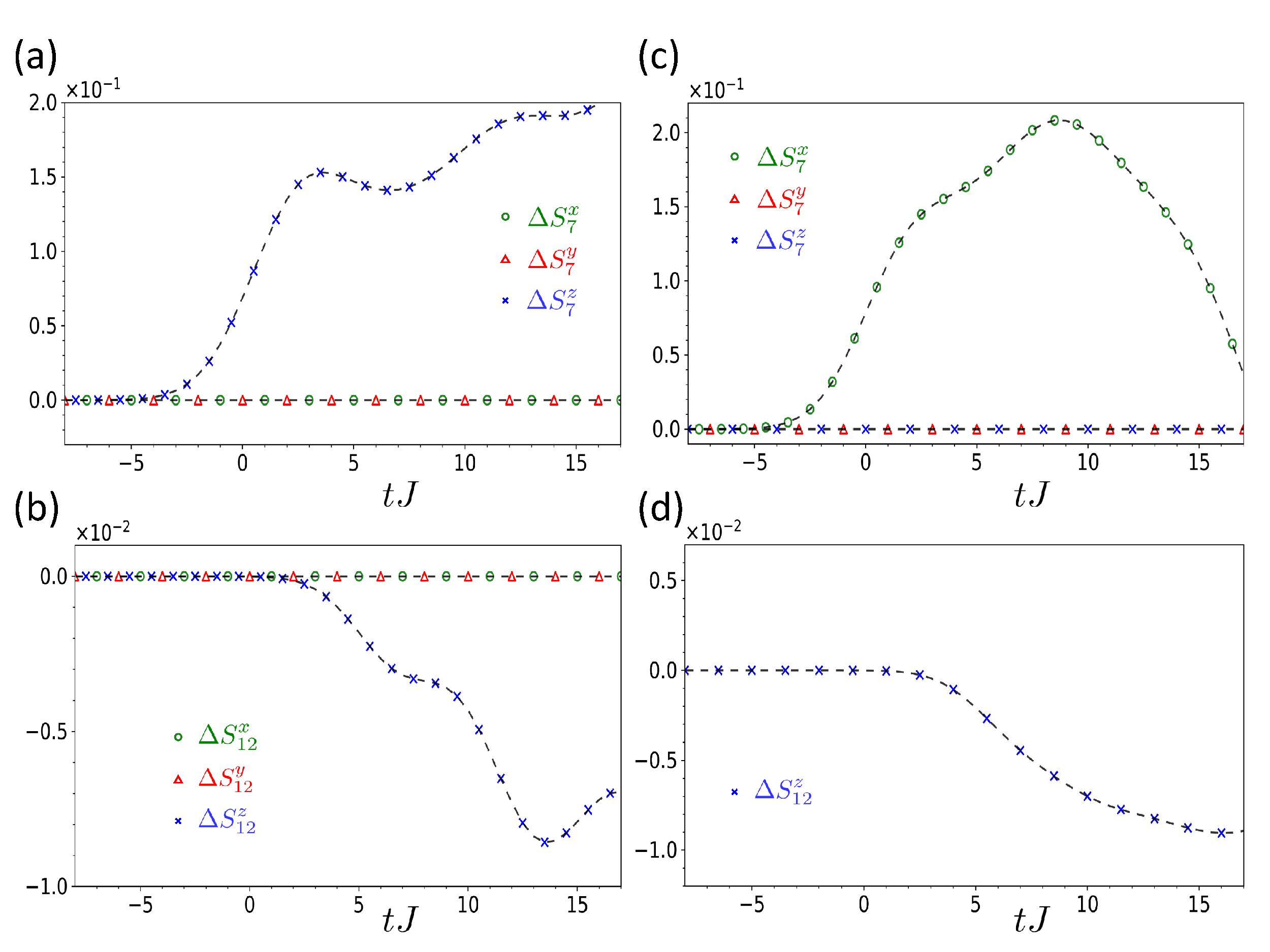}
  \caption{Real-time evolution of the Kitaev cluster with armchair edges.
    (a) and (b) represent the change in the spin moments in the L and R regions
    when the direction of the injected magnetic pulse is
    along the $S^z$ direction ${\bf n}={\bf z}$.
    (c) and (d) represent the change in the spin moments in the L and R regions
    when ${\bf n}={\bf x}$.
  }
  \label{fig:Arm}
\end{figure}
The spin oscillations in the Kitaev cluster with armchair edges
are shown in Fig.~\ref{fig:Arm}.
When the Gaussian magnetic pulse with ${\bf n}={\bf z}$ is introduced
in the L region at $t=0$,
the magnetic moments in the region L
are immediately induced in the $S^z$ direction,
as shown in Fig.~\ref{fig:Arm}(a).
In the middle region,
magnetic moments are never induced (not shown).
In the R region under the static magnetic field,
the finite spin oscillations appear in the $S^z$ direction,
as shown in Fig.~\ref{fig:Arm}(b).
When the magnetic pulse field is applied along the $S^x$ direction,
the magnetic moments are induced along the $S^x$ direction in the L region,
as shown in Fig.~\ref{fig:Arm}(c).
In the R region,
the induced magnetic moments have the components
in the $S^z$ direction, as shown in Fig.~\ref{fig:Arm}(d).
This fact is essentially the same as the result for the zigzag-edge Kitaev cluster
discussed above.
We have sometimes observed oscillation behavior in $S^x$ and $S^y$ directions (not shown).
We have also found that this phenomenon strongly depend
on the initial state in the Lanczos method,
while the oscillation in the $S^z$ direction never changes. 
This can be explained by the following.
There exists a local $Z_2$ operator at the edge:
$X=\sigma_{11}^y\sigma_{12}^z\sigma_6^z\sigma_5^x$
in the Kitaev system with the armchair edges [see Fig.~\ref{fig:system}(b)].
This operator commutes with the Hamiltonian with the finite field $h_R$,
which is similar to the operator $W_P$ in the middle region.
Therefore, one can prove that $\langle S^x_i \rangle=\langle S^y_i\rangle =0\;(i=6,12)$
for a certain eigenstate with $x=\pm 1$. 
However, in contrast to the fact that $w_P=1$ in the ground state,
the eigenvalue of $X$ is not uniquely determined,
meaning the existence of the ground state degeneracy originated from the edges.
  This leads to the initial state dependent oscillations
  in the $S^x$ and $S^y$ directions,
  while the oscillation in the $S^z$ direction parallel to the applied field
does not depend on the edge degeneracy.
  This can be clarified in the following.
  When the doubly degenerate normalized states $|\pm\rangle$ are defined such as
  $H|\pm\rangle=E_g|\pm\rangle$, $X|\pm\rangle=\pm |\pm\rangle$, and
  $\langle +|-\rangle=0$,
  the ground state is, in general, given as
  \begin{eqnarray}
    |\psi\rangle=\alpha |+\rangle+\beta |-\rangle,\label{eqp}
\end{eqnarray}
  where $|\alpha|^2+|\beta|^2=1$.
  Now, we furthermore introduce the operator $Y$,
  which satisfies $Y^2=1$, $\{X,Y\}=0$, $[Y,H]=0$,
  and $Y|\pm\rangle=e^{\pm i \theta}|\mp\rangle$ with a real constant $\theta$.
  One of the examples is
  $Y=\sigma_{12}^z\sigma_{11}^z\sigma_{10}^z\sigma_{9}^z\sigma_8^x\sigma_{14}^x\sigma_{13}^x\sigma_{19}^x$.
  In the case, we prove
  $\langle \psi|\sigma_{12}^z|\psi\rangle=\langle +|\sigma_{12}^z|+\rangle$
  since $\langle +|Y\sigma_{12}^z|+\rangle=0$ and $[\sigma_{12}^z,Y]=0$.
  Therefore, we can say that the $z$ component of the moment at the site 12
  does not depend on the initial state in the Lanczos method.
  On the other hand, the other components depends on $\alpha$ and $\beta$
  at $t=0$.
By these reasons, we conclude that the itinerant Majorana fermions propagate
the wave packet triggered by the magnetic pulse, while cannot propagate
its direction.

\section{Summary}

We have investigated the Majorana-mediated spin transport
in the Kitaev clusters with zigzag and armchair edges, 
applying the static magnetic field to one of the edges
and magnetic pulsed field to the other edges.
By means of the exact diagonalization methods,
the time-evolution of the systems has been examined.
In the Kitaev model with zigzag and armchair edges,
the spin oscillations in the R region appear along the $S^z$ direction
even when the magnetic pulse with ${\bf n}={\bf x}$
is applied to the L region.
Therefore, we conclude that the itinerant Majorana fermions propagate
the spin excitation injected by the magnetic pulse, while cannot propagate
its direction.
  The present results clarifying the spin propagation
  independent on field directions suggest that
  the information of a spin component is mainly stored by
  the localized Majorana fermions rather than itinerant ones.

\section*{Acknowledgements}
Parts of the numerical calculations are performed
in the supercomputing systems in ISSP, the University of Tokyo.
This work was supported by Grant-in-Aid for Scientific Research from
JSPS, KAKENHI Grant Nos.
JP17K05536, JP19H05821, JP21H01025, JP22K03525 (A.K.),
JP20K14412, JP21H05017 (Y.M.), JP19K03742, JP20H00122, JP22H01175 (J.N.),
JP21H01025 (H.T.), and 
JST CREST Grant Nos. JPMJCR1901 (Y.M.), JPMJCR17J5 (H.T.), and
JST PRESTO Grant No. JPMJPR19L5 (J.N.).

\end{document}